\documentstyle[12pt,amsfonts]{article}
\newcommand{\ehat}{ \hat U_{\epsilon} }

\newcommand{\define}{ \stackrel{\triangle}{=} }

\def\be{\begin{equation}}
\def\ee{\end{equation}}
\def\ba{\begin{array}}
\def\ea{\end{array}}

\def\d4{{\rm d}^4}

\begin{document}
%---------------------------------------------------
\title{\bf Unification of Electromagnetic Interactions and
                Gravitational Interactions }
\author{{Ning Wu}
\thanks{email address: wuning@mail.ihep.ac.cn}
\\
\\
{\small Institute of High Energy Physics, P.O.Box 918-1,
Beijing 100039, P.R.China}
\thanks{mailing address}}
\maketitle
\vskip 0.8in
%\noindent

~~

\vskip 0.8in
%\noindent

\begin{abstract}

Unified theory of gravitational interactions and electromagnetic
interactions is discussed in this paper. Based on gauge principle,
electromagnetic interactions and gravitational interactions are 
formulated in the same manner and are unified in a semi-direct product 
group of $U(1)$ Abel gauge group and gravitational gauge group.
\\

\end{abstract}
%-------------------------------------------------------

\newpage

\Roman{section}

\section{Introduction}

It is known that there are four kinds of fundamental interactions
in Nature, which are strong interactions, electromagnetic
interactions, weak interactions and gravitational interactions.
All these fundamental interactions can be described by gauge field
theories, which can be regarded as the common nature of all these
fundamental interactions. This provides us the possibilities to
unify different kinds of
fundamental interactions in the framework of gauge theory. The first
unification of fundamental interactions in human history is the
unification of electric interactions and magnetic interactions,
which is made by Maxwell in 1864. Now, we know that electromagnetic
theory is a $U(1)$ Able gauge theory. In 1921, H.Weyl tried to
unify electromagnetic interactions and gravitational interactions
in a theory which has local scale invariance\cite{1,2}. Weyl's
original work is not successful, however in his great attempt, he
introduced one of the most important concept in modern physics:
gauge transformation and gauge symmetry. After the foundation
of quantum mechanics, V.Fock, H.Weyl and W.Pauli found that quantum
electrodynamics is a $U(1)$ gauge invariant theory\cite{3,4,5}.
\\

Gauge treatment of gravity was suggested immediately after the
gauge theory birth itself\cite{29,30,31}. In the traditional gauge
treatment of gravity, Lorentz group is localized, and the
gravitational field is not represented by gauge
potential\cite{32,33,34}. It is represented by metric field. 
The theory has beautiful mathematical forms, but up to now, its
renormalizability is not proved. In other words,
it is conventionally considered to be non-renormalizable.
Recently, some new attempts were proposed to use Yang-Mills theory
to reformulate quantum gravity\cite{35,36,37,38}. In these new
approaches, the importance of gauge fields is emphasized. Some
physicists also try to use gauge potential to represent
gravitational field, some suggest that we should pay more
attention on translation group. Recently, Wu proposed a new quantum
gauge theory of gravity which is a renormalizable quantum
gravity\cite{39,40}. Based on gauge principle, space-time translation
group is selected to be the gravitational gauge group. After
localization of gravitational gauge group, the gravitational field
appears as the corresponding gauge potential. In this paper,
we use the spirit inspired by literature \cite{39,40} 
to discuss unification of fundamental interactions which is
based on gravitational gauge theory. Our main goal in this paper
is to discuss unification of gravitational interactions and 
electromagnetic interactions. \\

\section{Lagrangian}
\setcounter{equation}{0}

It is known that Quantum Electrodynamics(QED) is a $U(1)$ Able gauge
field theory. A general group element of $U(1)$ Able gauge group
is denoted as $U(x)$,
\be \label{2.1}
U(x) = e^{- i \alpha(x)}.
\ee
A general element of gravitational gauge group is denoted as 
$\ehat$\cite{39,40}
\be \label{2.2}
\ehat = E^{-i \epsilon^{\mu} \cdot \hat{P}_{\mu}}.
\ee
Because
\be \label{2.3}
(\hat{P}_{\mu} \alpha(x)) \not= 0,
\ee
$U(1)$ group element $U(x)$ does not commute with gravitational gauge
group element $\ehat$,
\be \label{2.4}
\lbrack U(x)~~,~~\ehat \rbrack \not= 0.
\ee
It means that all elements $g(x)$
\be \label{2.5}
g(x) = \ehat \cdot U(x)
\ee
form a semi-direct product group of $U(1)$ Able group and 
gravitational gauge group. We denote it as
\be \label{2.6}
GU(1) \define
U(1) \otimes_s Gravitational~Gauge~Group = \lbrace g(x)\rbrace.
\ee
This semi-direct product group is the symmetry group of the unified
theory of gravitational interactions and electromagnetic interactions.
\\

As an example, we discuss electromagnetic interactions and gravitational
interactions between Dirac field and electromagnetic field or gravitational
field. The traditional lagrangian density for electromagnetic interactions
between electromagnetic field $A_{\mu}$ and Dirac field $\psi$ is given by
\be \label{2.7}
- \frac{1}{4} \eta^{\mu \rho} \eta^{\nu \sigma}
A_{\mu \nu} A_{\rho \sigma}
- \bar{\psi}
( \gamma^{\mu} ( \partial_{\mu} - i e A_{\mu}  ) + m ) \psi,
\ee
where $A_{\mu \nu}$ is the field strength of electromagnetic field,
$e$ is the coupling constant of electromagnetic interactions and $m$ is
the mass of Dirac field. After considering gravitational interactions,
$A_{\mu \nu}$ is defined by
\be \label{2.8}
A_{\mu \nu} = (D_{\mu} A_{\nu}) - (D_{\nu} A_{\mu}).
\ee
In above relation, $D_{\mu}$ is the gravitational gauge covariant derivative,
\be \label{2.9}
D_{\mu} = \partial_{\mu} - i g C_{\mu} (x),
\ee
where $g$ is the gravitational coupling constant and
$C_{\mu}(x)$ is the gravitational gauge field which is a vector in gauge
group space\cite{39,40},
\be \label{2.10}
C_{\mu}(x) = C_{\mu}^{\alpha} (x) \hat{P}_{\alpha}.
\ee
The explicit expression for $A_{\mu\nu}$ is
\be \label{2.10a}
A_{\mu \nu} =  \partial_{\mu} A_{\nu} - \partial_{\nu} A_{\mu}
- g C_{\mu}^{\alpha} \partial_{\alpha} A_{\nu}
+ g C_{\nu}^{\alpha} \partial_{\alpha} A_{\mu}.
\ee
However, $A_{\mu \nu}$ is not a $U(1)$ gauge invariant field strength.
In order to define $U(1)$ gauge invariant field strength, we need a
matrix $G$. Define,
\be \label{2.11}
G \define (G_{\mu}^{\alpha}) 
\define ( \delta_{\mu}^{\alpha} - g C_{\mu}^{\alpha} ).
\ee
Its inverse matrix is denoted as $G^{-1}$,
\be \label{2.12}
G^{-1} = \frac{1}{I - gC} = (G^{-1 \mu}_{\alpha}).
\ee
They satisfy the following relations,
\be \label{2.13}
G_{\mu}^{\alpha} G^{-1 \nu}_{\alpha} = \delta_{\mu}^{\nu},
\ee
\be \label{2.14}
G_{\beta}^{-1 \mu} G^{ \alpha}_{\mu} = \delta_{\beta}^{\alpha}.
\ee
It can be proved that
\be \label{2.15}
D_{\mu}= G_{\mu}^{\alpha} \partial_{\alpha}.
\ee
The  field strength of gravitational gauge field is defined by
\be \label{2.16}
F_{\mu\nu} \define \frac{1}{-ig} \lbrack D_{\mu}~~,~~D_{\nu} \rbrack.
\ee
Its explicit expression is
\be \label{2.17}
F_{\mu\nu}(x) = \partial_{\mu} C_{\nu} (x)
-\partial_{\nu} C_{\mu} (x)
- i g C_{\mu} (x) C_{\nu}(x)
+ i g C_{\nu} (x) C_{\mu}(x).
\ee
$F_{\mu\nu}$ is also a vector in gauge group space,
\be \label{2.18}
F_{\mu\nu} (x) = F_{\mu\nu}^{\alpha}(x) \cdot \hat{P}_{\alpha},
\ee
where
\be \label{2.19}
F_{\mu\nu}^{\alpha} = \partial_{\mu} C_{\nu}^{\alpha}
-\partial_{\nu} C_{\mu}^{\alpha}
-  g C_{\mu}^{\beta} \partial_{\beta} C_{\nu}^{\alpha}
+  g C_{\nu}^{\beta} \partial_{\beta} C_{\mu}^{\alpha}.
\ee
The $U(1)$ gauge invariant field strength of electromagnetic field
is given by the following relation,
\be \label{2.20}
{\mathbb A}_{\mu \nu} = A_{\mu \nu}
+ g G^{-1 \lambda}_{\alpha} A_{\lambda} F_{\mu\nu}^{\alpha},
\ee
where $A_{\mu\nu}$ is defined by eq.(\ref{2.8}).
$GU(1)$ gauge covariant derivative
is
\be \label{2.21}
{\mathbb D}_{\mu} = \partial_{\mu}
- i e A_{\mu} - i g C_{\mu}.
\ee
\\

The gravitational gauge covariant and $U(1)$ gauge invariant
lagrangian density ${\cal L}_0$ is
\be \label{2.22}
{\cal L}_0 = - \bar{\psi}
( \gamma^{\mu}  {\mathbb D}_{\mu}  + m ) \psi
- \frac{1}{4} \eta^{\mu \rho} \eta^{\nu \sigma}
{\mathbb A}_{\mu \nu} {\mathbb A}_{\rho \sigma}
- \frac{1}{4} \eta^{\mu \rho} \eta^{\nu \sigma}
\eta_{2 \alpha \beta} F^{\alpha}_{\mu\nu}
F^{\beta}_{\rho\sigma},
\ee
where $\eta^{\mu\nu}$ is the Minkowski metric. The selection of
$\eta_{2\alpha\beta}$ is not unique\cite{39,40}. 
One selection is to set
\be
\eta_{2\alpha\beta} = \left(
\begin{array}{cccc}
-1 &0&0& 0 \\
0&1&0&0  \\
0&0&1&0  \\
0&0&0&1
\end{array}
\right).
\label{2.2201}
\ee
Another selection is to set
\be
\eta_{2\alpha \beta} = g_{\alpha \beta},
 \label{2.2202}
\ee
where
\be \label{2.27}
g_{\alpha \beta} \define \eta_{\mu \nu}
(G^{-1})_{\alpha}^{\mu} (G^{-1})_{\beta}^{\nu}.
\ee
The first choice gives out minimum model, so we use
the first choice in this paper.
\\
 
The full lagrangian
density ${\cal L}$ is defined by
\be \label{2.23}
{\cal L} = J(C) {\cal L}_0,
\ee
where $J(C)$ is a special factor to resume gravitational gauge symmetry
of the system. The selection of $J(C)$ is not unique. 
The simplest and most beautiful choice of $J(C)$ is
\cite{39,40},
\be \label{2.24}
J(C) = e^{I(C)},
\ee
where
\be \label{2.25}
I(C) = g \eta_{1\alpha}^{\mu} C_{\mu}^{\alpha}.
\ee
The definition of $\eta_{1\alpha}^{\mu}$ can be found in Ref.\cite{39,40}.
Another possible definition of $J(C)$ is
\be \label{2.26}
J(C) = \sqrt{- {\rm det} g_{\alpha \beta} },
\ee
where $ g_{\alpha \beta}$ is given by eq.(\ref{2.27})
In this paper, we will use definition eq.(\ref{2.24}), for this is the
simplest and most beautiful choice which has no poles in it. The action
of the system is
\be \label{2.28}
S = \int \d4 x{\cal L}.
\ee
\\

\section{Gauge Symmetry}
\setcounter{equation}{0}

Now, let's study the symmetry of the system. First, let's discuss
$U(1)$ gauge symmetry of the system. Under local $U(1)$ gauge
transformations, the transformations of various fields are
\be \label{2.29}
\psi \to \psi' = e^{-i \alpha} \psi,
\ee
\be \label{2.30}
C_{\mu} \to C'_{\mu} = C_{\mu},
\ee
\be \label{2.31}
C_{\mu}^{\alpha} \to C_{\mu}^{\prime\alpha} = C_{\mu}^{\alpha},
\ee
\be \label{2.32}
A_{\mu} \to A'_{\mu} = A_{\mu} - \frac{1}{e}
(D_{\mu} \alpha (x) ),
\ee
\be \label{2.33}
{\mathbb D}_{\mu} \to {\mathbb D}'_{\mu}
= e^{-i \alpha} {\mathbb D}_{\mu} e^{i \alpha} ,
\ee
\be \label{2.34}
A_{\mu\nu} \to A'_{\mu\nu} = A_{\mu\nu}
+ \frac{g}{e} F_{\mu\nu}^{\sigma} (\partial_{\sigma} \alpha(x)) ,
\ee
\be \label{2.35}
F^{\alpha}_{\mu\nu} \to F^{\prime\alpha}_{\mu\nu} = F^{\alpha}_{\mu\nu},
\ee
\be \label{2.36}
{\mathbb A}_{\mu\nu} \to {\mathbb A}'_{\mu\nu} = {\mathbb A}_{\mu\nu}.
\ee
From above transformation relations, we could see that gravitational gauge
field keeps unchanged under $U(1)$ Abel gauge transformations, and
$A_{\mu\nu}$ is not a $U(1)$ gauge invariant field strength while
${\mathbb A}_{\mu\nu}$ is a $U(1)$ gauge invariant. Under $U(1)$ gauge
transformation, lagrangian density ${\cal L}_0$ is invariant,
\be \label{2.37}
{\cal L}_0 \to {\cal L}'_0 = {\cal L}_0.
\ee
Because $J(C)$ is also $U(1)$ gauge invariant, the full lagrangian density
${\cal L}$ and action $S$ of the system are also $U(1)$ gauge invariant,
\be \label{2.38}
{\cal L} \to {\cal L}' = {\cal L},
\ee
\be \label{2.39}
S \to S' = S.
\ee
Therefore, the system has local $U(1)$ gauge symmetry. \\

Under gravitational gauge transformations, various fields and operators
transform as
\be \label{2.40}
\psi \to \psi' = (\ehat \psi) ,
\ee
\be \label{2.41}
C_{\mu} \to C'_{\mu} = \ehat C_{\mu} \ehat^{-1}
- \frac{1}{ig} \ehat (\partial_{\mu} \ehat^{-1} ),
\ee
\be \label{2.42}
G_{\mu}^{\alpha} \to G_{\mu}^{\prime\alpha}
= \Lambda^{\alpha}_{~\beta} \ehat G_{\mu}^{\beta} \ehat^{-1},
\ee
\be \label{2.43}
G^{-1\mu}_{\alpha} \to G^{\prime -1 \mu}_{\alpha}
= \Lambda_{\alpha}^{~\beta} \ehat G^{-1\mu}_{\beta} \ehat^{-1},
\ee
\be \label{2.4301}
g_{\alpha\beta} \to g^{\prime}_{\alpha\beta}
= \Lambda_{\alpha}^{~\alpha_1} \Lambda_{\beta}^{~\beta_1} 
\ehat g_{\alpha\beta} \ehat^{-1},
\ee
\be \label{2.44}
A_{\mu} \to A'_{\mu} = \ehat A_{\mu} \ehat^{-1},
\ee
\be \label{2.45}
{\mathbb D}_{\mu} \to {\mathbb D}'_{\mu}
 = \ehat {\mathbb D}_{\mu} \ehat^{-1},
\ee
\be \label{2.46}
A_{\mu\nu} \to A'_{\mu\nu} = \ehat A_{\mu\nu}\ehat^{-1} ,
\ee
\be \label{2.47}
F^{\alpha}_{\mu\nu} \to F^{\prime\alpha}_{\mu\nu}
= \Lambda^{\alpha}_{~\beta} \ehat F^{\beta}_{\mu\nu} \ehat^{-1},
\ee
\be \label{2.48}
{\mathbb A}_{\mu\nu} \to {\mathbb A}'_{\mu\nu}
= \ehat {\mathbb A}_{\mu\nu} \ehat^{-1},
\ee
\be \label{2.49}
J(C) \to J'(C') = J \cdot \ehat J(C) \ehat^{-1},
\ee
where $J$ is the Jacobian of the corresponding transformation
which is given by
\be \label{2.50}
J = det \left(\frac{\partial (x - \epsilon)^{\mu}}
{\partial x^{\nu}} \right).
\ee
Using all these relations, we can prove that the lagrangian density
${\cal L}_0$ is gravitational gauge covariant,
\be \label{2.51}
{\cal L}_0 \to {\cal L}'_0 = \ehat {\cal L}_0 \ehat^{-1}.
\ee
Then eq.(\ref{2.49}), eq.(\ref{2.50}) and eq.(\ref{2.23}) give out
\be \label{2.52}
{\cal L} \to {\cal L}' =J \cdot \ehat {\cal L} \ehat^{-1}.
\ee
Using the following relation,
\be \label{2.53}
\int {\rm d}^4 x J(\ehat f(x)) = \int {\rm d}^4 x f(x),
\ee
we can prove that the action of the system is gravitational gauge
invariant,
\be \label{2.54}
S \to S' = S.
\ee
Therefore, the system has local gravitational gauge symmetry.\\

The action defined by eq.(\ref{2.28}) has both local $U(1)$ gauge
symmetry and local gravitational gauge symmetry. It means that it
has local $GU(1)$ gauge symmetry. The $GU(1)$ gauge
transformations of various fields and operators are
\be \label{2.55}
\psi \to \psi' = (g(x) \psi) ,
\ee
\be \label{2.56}
C_{\mu} \to C'_{\mu} = \ehat C_{\mu} \ehat^{-1}
- \frac{1}{ig} \ehat (\partial_{\mu} \ehat^{-1} ),
\ee
\be \label{2.57}
G_{\mu}^{\alpha} \to G_{\mu}^{\prime\alpha}
= \Lambda^{\alpha}_{~\beta} ~~g(x) G_{\mu}^{\beta} g^{-1}(x),
\ee
\be \label{2.58}
G^{-1\mu}_{\alpha} \to G^{\prime -1 \mu}_{\alpha}
= \Lambda_{\alpha}^{~\beta}~~ g(x) G^{-1\mu}_{\beta} g^{-1}(x),
\ee
\be \label{2.59}
A_{\mu} \to A'_{\mu} = g(x) \left\lbrack A_{\mu} - \frac{1}{e}
(D_{\mu} \alpha (x) ) \right\rbrack g^{-1}(x),
\ee
\be \label{2.60}
{\mathbb D}_{\mu} \to {\mathbb D}'_{\mu}
 = g(x) {\mathbb D}_{\mu} g^{-1}(x),
\ee
\be \label{2.61}
A_{\mu\nu} \to A'_{\mu\nu} = g(x) \left\lbrack A_{\mu\nu}
+ \frac{g}{e} F_{\mu\nu}^{\sigma} (\partial_{\sigma} \alpha(x))
\right\rbrack g^{-1}(x),
\ee
\be \label{2.62}
F^{\alpha}_{\mu\nu} \to F^{\prime\alpha}_{\mu\nu}
= \Lambda^{\alpha}_{~\beta}~ ~g(x) F^{\beta}_{\mu\nu} g^{-1}(x),
\ee
\be \label{2.63}
{\mathbb A}_{\mu\nu} \to {\mathbb A}'_{\mu\nu}
= g(x) {\mathbb A}_{\mu\nu} g^{-1}(x),
\ee
\be \label{2.64}
J(C) \to J'(C') = J \cdot g(x) J(C) g^{-1}(x),
\ee
where $g(x)$ is given by eq.(\ref{2.5}). The action eq.(\ref{2.28})
is invariant under these transformations.  \\

\section{Interactions}
\setcounter{equation}{0}

Define
\be \label{2.65}
A_{0\mu\nu} = \partial_{\mu} A_{\nu}
-\partial_{\nu} A_{\mu},
\ee
\be \label{2.66}
F_{0\mu\nu}^{\alpha} = \partial_{\mu} C_{\nu}^{\alpha}
-\partial_{\nu} C_{\mu}^{\alpha}.
\ee
Then, we can separate the lagrangian density eq.(\ref{2.23}) into
free lagrangian density ${\cal L}_F$ and interaction lagrangian
density ${\cal L}_I$,
\be \label{2.67}
{\cal L} = {\cal L}_F + {\cal L}_I,
\ee
where,
\be \label{2.68}
{\cal L}_F = - \frac{1}{4} \eta^{\mu \rho} \eta^{\nu
\sigma} A_{0 \mu \nu} A_{0 \rho \sigma} - \bar{\psi} (
\gamma^{\mu}
\partial_{\mu}  + m ) \psi - \frac{1}{4} \eta^{\mu \rho} \eta^{\nu
\sigma} \eta_{2 \alpha \beta } F_{0 \mu \nu}^{\alpha} F_{0 \rho
\sigma}^{\beta},
\ee
\be \label{2.69}
\begin{array}{rcl}
{\cal L}_I &=& - \frac{1}{4} \eta^{\mu \rho} \eta^{\nu \sigma}
(\sum_{n=1}^{\infty} \frac{1}{n!}
( g \eta_{1 \alpha_1}^{\mu_1} C_{\mu_1}^{\alpha_1} )^n )
A_{0 \mu \nu} A_{0 \rho \sigma}\\
&&\\
&& - (\sum_{n=1}^{\infty} \frac{1}{n!}
( g \eta_{1 \alpha_1}^{\mu_1} C_{\mu_1}^{\alpha_1} )^n )
\bar{\psi} ( \gamma^{\mu} \partial_{\mu}  + m ) \psi \\
&&\\
&& - \frac{1}{4} \eta^{\mu \rho} \eta^{\nu\sigma}
\eta_{2 \alpha \beta } (\sum_{n=1}^{\infty} \frac{1}{n!}
( g \eta_{1 \alpha_1}^{\mu_1} C_{\mu_1}^{\alpha_1} )^n )
F_{0 \mu \nu}^{\alpha} F_{0 \rho\sigma}^{\beta} \\
&&\\
&& + i e \cdot e^{I(C)} \bar{\psi} \gamma^{\mu} \psi A_{\mu}
+ g e^{I(C)} \bar\psi \gamma^{\mu} (\partial_{\alpha} \psi)
C_{\mu}^{\alpha}\\
&&\\
&& + g e^{I(C)} \eta^{\mu \rho} \eta^{\nu \sigma}
(\partial_{\mu} A_{\nu} -\partial_{\nu} A_{\mu})
 C_{\rho}^{\alpha} (\partial_{\alpha} A_{\sigma}) \\
&&\\
&&- \frac{g}{2} e^{I(C)} \eta^{\mu \rho} \eta^{\nu \sigma} A_{\mu
\nu} G^{-1 \lambda }_{\alpha} A_{\lambda} F_{\rho \sigma}^{\alpha}\\
&&\\
&& -\frac{g^2}{4} e^{I(C)} \eta^{\mu \rho} \eta^{\nu \sigma} G^{-1
\kappa}_{\alpha} G^{-1 \lambda }_{\beta} A_{\kappa} A_{\lambda}
F_{\mu \nu}^{\alpha} F_{\rho \sigma}^{\beta} \\
&&\\
&& - \frac{g^2}{2} e^{I(C)} \eta^{\mu \rho} \eta^{\nu \sigma}
( C_{\mu}^{\alpha} C_{\rho}^{\beta} (\partial_{\alpha} A_{\nu} )
(\partial_{\beta} A_{\sigma} )
- C_{\nu}^{\alpha} C_{\rho}^{\beta} (\partial_{\alpha} A_{\mu} )
(\partial_{\beta} A_{\sigma} )  )  \\
&&\\
&&  + g e^{I(C)} \eta^{\mu \rho} \eta^{\nu \sigma}
\eta_{2 \alpha \beta} (\partial_{\mu} C_{\nu }^{\alpha}
- \partial_{\nu} C_{\mu}^{\alpha})
C_{\rho}^{\delta} (\partial_{\delta} C_{\sigma}^{\beta}) \\
&&\\
&&  - \frac{1}{2} g^2 e^{I(C)} \eta^{\mu \rho} \eta^{\nu \sigma}
\eta_{2 \alpha \beta}
(C_{\mu}^{\delta} \partial_{\delta} C_{\nu}^{\alpha}
- C_{\nu}^{\delta} \partial_{\delta} C_{\mu}^{\alpha} )
C_{\rho}^{\epsilon} (\partial_{\epsilon} C_{\sigma}^{\beta}).
\end{array}
\ee
From eq.(\ref{2.68}), we can write out propagators of gravitational
gauge field, electromagnetic field and Dirac field. 
From the interaction lagrangian ${\cal L}_I$, we can see that 
Dirac field, electromagnetic field and gravitational gauge field 
couple each other. From eq.(\ref{2.69}), we can write out 
Feynman Rules for various interaction vertices which is useful
for calculation of Feynman diagrams. \\

\section{Equations of Motion and Conserved Currents}
\setcounter{equation}{0}

The Euler-Lagrangian equation of motion of Dirac field is
\be \label{2.71}
\left\lbrack \gamma^{\mu} ( \partial_{\mu} - ie A_{\mu}
- g C_{\mu}^{\alpha} \partial_{\alpha} ) + m \right\rbrack
\psi = 0.
\ee
The equation of motion of electromagnetic field is
\be \label{2.72}
\ba{rcl}
\partial^{\mu} {\mathbb A}_{\mu\nu} &=&
- i e \bar\psi \gamma_{\nu} \psi
+ g \eta^{\lambda\rho} \partial_{\mu}
(C_{\lambda}^{\mu} {\mathbb A}_{\rho\nu} )\\
&&\\
&& -g \eta^{\kappa\rho} G_{\kappa}^{\mu} {\mathbb A}_{\rho\nu}
\partial_{\mu} ( \eta^{\lambda}_{1\alpha} C_{\lambda}^{\alpha} )
+ \frac{g}{2} \eta^{\mu_1 \rho} \eta^{\nu_1 \sigma}
\eta_{\nu \lambda} G^{-1 \lambda}_{\alpha}
F^{\alpha}_{\mu_1 \nu_1} {\mathbb A}_{\rho\sigma}.
\ea
\ee
The equation of motion of gravitational gauge field is
\be \label{2.73}
\partial^{\mu}(\eta^{\nu\sigma} \eta_{2 \alpha\beta} 
F_{\mu\sigma}^{\beta} )
= - g T_{g\alpha}^{\nu},
\ee
where $T_{g\beta}^{\sigma}$ is the gravitational energy-momentum
tensor, whose explicit expression is
\be \label{2.74}
\ba{rcl}
T_{g \alpha}^{\nu} &=& \bar\psi \gamma^{\nu} \partial_{\alpha} \psi
- \eta^{\mu_1 \rho} \eta^{\nu \sigma} (\partial_{\alpha} A_{\mu_1})
{\mathbb A}_{\rho\sigma}
 - \eta^{\mu_1 \rho} \eta^{\nu \sigma} \eta_{2 \alpha_1 \beta}
F_{\rho\sigma}^{\beta} (\partial_{\alpha} C_{\mu_1}^{\alpha_1})
+ \eta^{\nu}_{1\alpha} {\cal L}_0  \\
&&\\
&& -\eta^{\mu_1 \rho} \eta^{\nu \sigma} \eta_{2\alpha\beta}
(\partial_{\mu} ({C_{\mu_1}^{\mu} F_{\rho\sigma}^{\beta}}))
+ \eta^{\mu_1 \rho} \eta^{\nu \sigma}
\partial_{\mu} ( G^{-1 \lambda}_{\alpha} G^{\mu}_{\mu_1}
A_{\lambda} {\mathbb A}_{\rho\sigma} ) \\
&&\\
&& + \eta^{\mu \rho} \eta^{\nu \sigma} \eta_{2\alpha\beta}
( D_{\mu} (\eta^{\lambda}_{1\alpha_1} C_{\lambda}^{\alpha_1} ))
F_{\rho\sigma}^{\beta}
- \frac{g}{2} \eta^{\mu_1 \rho} \eta^{\nu_1 \sigma}
G^{-1\nu}_{\alpha_1} G^{-1\lambda}_{\alpha} A_{\lambda}
{\mathbb A}_{\rho\sigma} F_{\mu_1\nu_1}^{\alpha_1} \\
&&\\
&& - g \eta^{\mu_1 \rho} \eta^{\nu \sigma}
G^{-1 \lambda}_{\alpha_1} A_{\lambda} {\mathbb A}_{\rho\sigma}
( \partial_{\alpha} C_{\mu_1}^{\alpha_1} )
+ g \eta^{\mu \rho} \eta^{\nu \sigma}
( D_{\mu} (\eta^{\lambda_1}_{1 \alpha_1} C_{\lambda_1}^{\alpha_1}))
G^{-1\lambda}_{\alpha} A_{\lambda} {\mathbb A}_{\rho\sigma}.
\ea
\ee
\\

The global gravitational gauge symmetry gives out inertial
energy-momentum tensor $T_{i \alpha}^{\mu}$,
\be \label{2.75}
\ba{rcl}
T_{i \alpha}^{\mu} &=& e^{I(C)} \lbrack
\bar\psi \gamma^{\nu} G_{\nu}^{\mu} \partial_{\alpha} \psi
+ \eta^{\mu_1 \rho} \eta^{\nu\sigma}
G_{\mu_1}^{\mu} {\mathbb A}_{\rho\sigma} (\partial_{\alpha} A_{\nu})\\
&&\\
&& + \eta^{\mu \rho} \eta^{\nu\sigma} \eta_{2\beta\gamma}
F_{\rho\sigma}^{\gamma} (\partial_{\alpha} C_{\nu}^{\beta} )
- g \eta^{\mu_1 \rho} \eta^{\nu\sigma} \eta_{2\beta\gamma}
C_{\mu_1}^{\mu} F_{\rho\sigma}^{\gamma}
(\partial_{\alpha} C_{\nu}^{\beta}) \\
&&\\
&& +g \eta^{\mu_1 \rho} \eta^{\nu\sigma} G^{-1 \lambda}_{\beta}
G^{\mu}_{\mu_1} A_{\lambda} {\mathbb A}_{\rho\sigma}
(\partial_{\alpha} C_{\nu}^{\beta})
+ \delta_{\alpha}^{\mu} {\cal L}_0
\rbrack.
\ea
\ee
Compare eq.(\ref{2.74}) with eq.(\ref{2.75}), we can easy find that
gravitational energy-momentum tensor is not equivalent to the
inertial energy-momentum tensor. But when gravitational gauge field
vanishes, they are equivalent. In this case, they are
\be \label{2.76}
T_{i \alpha}^{\mu} = T_{g \alpha}^{\mu} =
\bar\psi \gamma^{\mu} \partial_{\alpha} \psi
+ \eta^{\mu\rho} \eta^{\nu\sigma}
A_{\rho\sigma} ( \partial_{\alpha} A_{\nu} )
+ \delta^{\mu}_{\alpha} {\cal L}_0,
\ee
which is what we expected in traditional quantum field theory. \\

If we denote
\be \label{2.77}
\ba{rcl}
{\mathbb J}_{\nu} &=&
 i  \bar\psi \gamma_{\nu} \psi
- \frac{g}{e} \eta^{\lambda\rho} \partial_{\mu}
(C_{\lambda}^{\mu} {\mathbb A}_{\rho\nu} )
 + \frac{g}{e} \eta^{\kappa\rho} G_{\kappa}^{\mu} {\mathbb A}_{\rho\nu}
\partial_{\mu} ( \eta^{\lambda}_{1\alpha} C_{\lambda}^{\alpha} )\\
&&\\
&&- \frac{g}{2e} \eta^{\mu_1 \rho} \eta^{\nu_1 \sigma}
\eta_{\nu \lambda} G^{-1 \lambda}_{\alpha}
F^{\alpha}_{\mu_1 \nu_1} {\mathbb A}_{\rho\sigma}.
\ea
\ee
Then, eq.(\ref{2.72}) can be changed into
\be \label{2.78}
\partial^{\mu} {\mathbb A}_{\mu\nu} = -e {\mathbb J}_{\nu}.
\ee
Because,
\be \label{2.79}
\partial^{\nu} \partial^{\mu} {\mathbb A}_{\mu\nu} = 0,
\ee
${\mathbb J}_{\nu}$ is a conserved current,
\be \label{2.80}
\partial^{\nu} {\mathbb J}_{\nu} = 0.
\ee
It means that $\bar\psi \gamma_{\nu} \psi$ is no longer a conserve
current,
\be \label{2.81}
\partial^{\nu} (\bar\psi \gamma_{\nu} \psi)  \not= 0.
\ee
Both gravitational field $C_{\mu}^{\alpha}$ and electromagnetic field
$A_{\mu}$ contribute some to generalized electromagnetic current. In other
words, in the unified theory, electromagnetic field is also
a source of itself, which originates from the non-Able nature of
the semi-direct product group $GU(1)$.
But if gravitational field vanishes, the electromagnetic current
${\mathbb J}_{\mu}$ will return to
\be \label{2.82}
{\mathbb J}_{\nu} = i \bar\psi \gamma_{\nu} \psi,
\ee
which is just the traditional electromagnetic current.
\\

\section{Summary}

In this paper, gravitational interactions and electromagnetic
interactions are unified in a semi-direct product group.
Because generator of $U(1)$ group and generators 
of gravitational gauge group have different dimension, 
that is, generator of $U(1)$ group are dimensionless 
while generators of gravitational gauge group have negative
mass dimension, it is hard to unify $U(1)$ gauge
interactions and gravitational gauge interactions in a simple
group. Because of the difference of dimensions of 
generators, we need at least two independent parameters for
coupling constant in any kind of unified theory. When
we unify $U(1)$ gauge interactions and gravitational gauge
interactions in $GU(1)$ group, we only need two independent
parameters for coupling constant, this unified theory can
be regarded as a kind of minimal theory of unification. 
It is impossible to unify gravitational interactions and 
electromagnetic interactions in a simple group in which
only one independent coupling constant is used. \\

Because $U(1)$ gauge group and gravitational gauge group
are unified in a semi-direct product group, not in a 
direct product group,
field strength of gravitational gauge field joins into
the definition of gauge covariant field strength of
$U(1)$ gauge field. This will cause additional interactions
between $U(1)$ gauge fields and gravitational gauge field
and affect the definition of generalized electromagnetic
current. 
\\

\end{document}